\documentclass[a4paper]{article}

\usepackage{INTERSPEECH2019}
\usepackage{mathtools}
\usepackage[percent]{overpic}
\usepackage{multirow}
\usepackage{array}
\usepackage[keeplastbox]{flushend}
\DeclarePairedDelimiter{\floor}{\lfloor}{\rfloor}

\title{Keyword Spotting for Hearing Assistive Devices Robust to External Speakers}
\name{Iv\'an L\'opez-Espejo$^1$, Zheng-Hua Tan$^1$ and Jesper Jensen$^{1,2}$}
\address{
  $^1$Department of Electronic Systems, Aalborg University, Denmark\\
  $^2$Oticon A/S, Denmark}
\email{\texttt{\{ivl,zt,jje\}@es.aau.dk}, \texttt{jesj@oticon.com}}

\begin{document}

\maketitle
\begin{abstract}
Keyword spotting (KWS) is experiencing an upswing due to the pervasiveness of small electronic devices that allow interaction with them via speech. Often, KWS systems are speaker-independent, which means that any person ---user or not--- might trigger them. For applications like KWS for hearing assistive devices this is unacceptable, as only the user must be allowed to handle them. In this paper we propose KWS for hearing assistive devices that is robust to external speakers. A state-of-the-art deep residual network for small-footprint KWS is regarded as a basis to build upon. By following a multi-task learning scheme, this system is extended to jointly perform KWS and users' own-voice/external speaker detection with a negligible increase in the number of parameters. For experiments, we generate from the Google Speech Commands Dataset a speech corpus emulating hearing aids as a capturing device. Our results show that this multi-task deep residual network is able to achieve a KWS accuracy relative improvement of around 32\% with respect to a system that does not deal with external speakers.
\end{abstract}

\section{Introduction}
\label{sec:intro}

Keyword spotting (KWS) aims at detecting a series of words from an audio stream comprising speech. This technology has become a popular research topic as it is considered a keystone for voice-based activation of virtual assistants (e.g., smart speakers) by means of keywords or wake-up-words \cite{Hoy18}.

Similarly, KWS may allow a hearing impaired person to initiate certain actions on her/his hearing assistive device, e.g., raising or lowering the volume. Since hearing aids are low computational resource devices, it is crucial that the KWS systems deployed on them have a small footprint (i.e., low memory and computational complexity).

Over the recent period, small-footprint KWS has attracted the attention of speech researchers due to the rapid development of deep learning \cite{Arik17}. For instance, in \cite{Parada15b}, Chen \emph{et al.} proposed extracting keyword embeddings from a word-based long short-term memory (LSTM) acoustic model. During testing, successive embeddings are extracted from a sliding window and compared against keyword templates. This approach is shown to outperform a KWS system based on phoneme posteriorgram with dynamic time warping (DTW) \cite{Hazen09}.

Working towards end-to-end models is the trend in the context of small-footprint KWS. In \cite{Parada14}, a deep neural network (DNN) is trained to predict keywords followed by a posterior handling technique outputting a confidence score. Robustness to both background noise and far-field conditions of this DNN-based KWS system is further improved in \cite{Parada15} through a combination of multi-style training and automatic gain control. Growing from these works, the use of convolutional neural networks (CNNs) is explored in \cite{Parada15c}, which allows for improving DNN performance with far fewer parameters. This is a remarkable finding since maximizing KWS accuracy while having a compact model is essential in the context of low-resource devices. In this respect, it is not surprising that the number of both multiplications and parameters is directly correlated with the energy usage of the device that the KWS model is running on \cite{Tang18}. Finally, very recent work \cite{Tang18b} investigates deep residual learning and dilated convolutions for small-footprint KWS and outperforms the previous state-of-the-art CNN-based system of \cite{Parada15c} on the Google Speech Commands Dataset \cite{Warden18}.

To the best of our knowledge, all of the above works are developed to be speaker-independent. This means that anyone should be able to trigger the proposed KWS systems. Nevertheless, for a number of applications we may want that only a particular speaker can interact with the system. This is often the case for hearing assistive devices, e.g., hearing aid systems, where the user must be the only one allowed to trigger actions on her/his hearing assistive device. Therefore, in this paper we explore KWS for hearing assistive devices that is robust to external speakers. Drawing from the deep residual network of \cite{Tang18b}, we consider multi-task learning to jointly perform KWS and own-voice/external speaker detection with a negligible increase in the number of parameters. For experimental purposes, a speech database emulating hearing aids as a capturing device is created from the Google Speech Commands Dataset. Our experimental results show that, thanks to exploiting the hearing aid multi-microphone signals, this multi-task deep residual network is able to improve KWS accuracy by about 32\% compared to a system that does not deal with external speakers.


\section{Deep Residual Learning for KWS}
\label{sec:kws}

\begin{figure*}[t]
  \centering
  \includegraphics[width=0.95\linewidth]{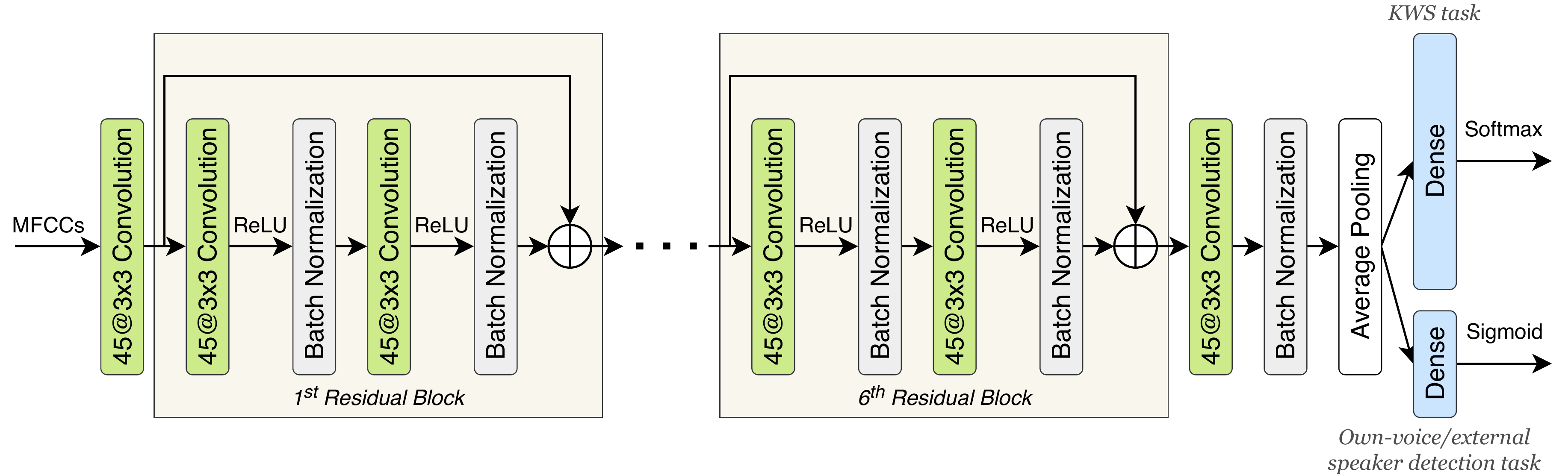}
  \caption{Diagram of the multi-task deep residual network for KWS with own-voice/external speaker detection.}
  \label{fig:resnet}
\end{figure*}

In this section we briefly review the deep residual network for small-footprint KWS proposed in \cite{Tang18b} (\texttt{res15}), as this is regarded as a basis to build upon. This architecture is based on the work of He \emph{et al.} \cite{He16}, where the authors proposed residual learning to tackle the performance degradation that occurs when CNNs are too deep. Let $\mathbf{x}_{l-1}$ be the input to a particular layer $l$. For too deep networks, it might be easier to optimize the residual mapping $\mathcal{H}_{l}^{l+k}(\mathbf{x}_{l-1})=\mathcal{F}_{l}^{l+k}(\mathbf{x}_{l-1})+\mathbf{x}_{l-1}$ between layers $l$ and $l+k$ ($k\in\mathbb{N}$) than the original mapping $\mathcal{F}_{l}^{l+k}(\mathbf{x}_{l-1})$ \cite{He16}. The above residual mapping can be simply accomplished by means of identity shortcut connections (identity mapping), i.e., those that skip $k+1$ layers.

The \texttt{res15} architecture \cite{Tang18b}, which uses Mel-frequency cepstral coefficients (MFCCs) as input, may be described as follows (see left part of Figure \ref{fig:resnet}). The first layer is a convolutional layer, after which there is a total of six residual blocks with identity mapping. Every residual block comprises two additional convolutional layers each of them followed by a rectified linear unit (ReLU) activation function and a batch normalization layer. Convolutional layers in the residual blocks apply dilated convolutions with a dilation rate of $\left(2^{\floor*{\frac{l}{3}}},2^{\floor*{\frac{l}{3}}}\right)$, where $l=0,...,11$ refers to the successive layers of this type and $\floor*{\cdot}$ denotes the floor function. Then, a non-residual convolutional layer with $(16,16)$ convolution dilation, another batch normalization layer and an average pooling layer are appended to the deepest residual block. Finally, a fully-connected (dense) layer with softmax activation is used for keyword classification.

It should be noticed that for all the convolutional layers the bias vector is zero, the kernel size is $3\times 3$ and the number of feature maps is set to $45$. The reader is referred to \cite{Tang18b} for further details on this deep residual network.

\section{Multi-task Learning for KWS and Own-Voice/External Speaker Detection}
\label{sec:multitask}

We employ the state-of-the-art \texttt{res15} described in the previous section to perform KWS on hearing assistive devices. Let $\mathbf{X}$ be the input speech features to the model. In order to also let the deep residual network detect whether the user, $S_u$, or an external speaker, $S_e$, is trying to trigger the KWS system, we extend it with an additional output providing an estimate of the conditional probability $P(S_u|\mathbf{X})=1-P(S_e|\mathbf{X})$. Then, keyword prediction from $\mathbf{X}$ is considered only if the estimate of $P(S_u|\mathbf{X})$ is above a certain threshold, e.g., 0.5.

When multi-task learning is considered for rather heterogeneous (i.e., dissimilar) tasks (as is our case), it makes sense that the task-specific output layers depend on different neuron activations. It is worth noticing that we conducted preliminary experiments by alternatively appending an average pooling layer and the own-voice/external speaker detection layer to the output of each residual block. However, these experiments revealed that there are no statistically significant differences in terms of both KWS and own-voice/external speaker detection accuracies as a function of the placement of the own-voice/external speaker detection layer. Therefore, in order to reduce the number of multiplications, we end up with the very simple solution to append such a layer, which consists of a fully-connected layer with one neuron and sigmoid activation, to the already existing average pooling layer, as can be seen in Figure \ref{fig:resnet}.

We will show in Section \ref{sec:results} that this multi-task approach is highly effective to jointly perform KWS and own-voice/external speaker detection. We will also show that combining the two network outputs leads to significantly improved KWS scores.

\subsection{Input Features}

First, the input audio signal is filtered by a band-pass filter with low and high cut-off frequencies of 20 Hz and 4 kHz, respectively \cite{Tang18b}. Then, the filtered signal is split into frames using a 30 ms Hann window with a 10 ms shift. Finally, 40 MFCCs are computed from each time frame and the resulting two-dimensional matrix is, after mean and standard deviation normalization, the input to the model \cite{Tang18b}.

In case of a multi-microphone signal (as the one from a hearing assistive device), we apply the above procedure independently to each channel and the resulting MFCC matrices are stacked across the quefrency dimension. While the number of parameters of the multi-task deep residual network remains the same, the number of multiplications experiences a relative increase of approximately $105\times(N-1)\%$, where $N$ is the number of microphones.

\subsection{Loss Weight Selection}

To dynamically prioritize the most difficult task during training we experimented with dynamic task prioritization \cite{Guo18}. Let us denote the KWS and own-voice/external speaker detection tasks as $T_1$ and $T_2$, respectively, and let $\mathcal{T}=\{T_1,T_2\}$ designate the total set of tasks. Furthermore, let $\lambda_j(i)$ and $\mathcal{L}_j(i)$ denote the loss weight and training loss, respectively, for the $j$-th task at epoch $i$. The total loss at the $i$-th epoch can be expressed as
\begin{equation}
 \mathcal{L}(i)=\sum_{j=1}^{|\mathcal{T}|}\lambda_j(i)\mathcal{L}_j(i),
\end{equation}
where $|\cdot|$ means cardinality. In \cite{Guo18}, the loss weights are updated according to
\begin{equation}
 \lambda_j(i)=-(1-\bar{\kappa}_j(i))\log(\bar{\kappa}_j(i)),
 \label{eq:weight}
\end{equation}
where $\bar{\kappa}_j(i)$ is the training accuracy resulting from an exponential moving average with forgetting factor $\alpha=0.75$. Then, the loss weights are normalized across tasks in such a way that $\sum_j\lambda_j(i)=|\mathcal{T}|\;$ $\forall i$. Thus, the higher the training accuracy for a given task, the lower its loss weight.

We also tested two variants of dynamic task prioritization. The first one simply consisted of using $\lambda_j(i)=\bar{\kappa}_j^{-1}(i)$ instead of (\ref{eq:weight}). Inspired by \cite{Chen18}, the second variant considered the training loss instead of the training accuracy. Let $\bar{\mathcal{L}}_j(i)$ be the exponential moving average of $\mathcal{L}_j(i)$. The loss weights are updated with the loss ratio $\bar{\mathcal{L}}_j(i)/\mathcal{L}_j(0)$, where $\mathcal{L}_j(0)=\log(C_j)$ is a theoretical initial cross-entropy loss across $C_j$ classes. This loss ratio can be understood as a measure of the inverse training rate of task $T_j$ \cite{Chen18}. Finally, we again normalize the loss weights across tasks as described above. In this case, the lower the loss ratio for a given task, the lower its loss weight.

Our preliminary experiments showed that there are no statistically significant differences in terms of both KWS and own-voice/external speaker detection accuracies with respect to using constant equal loss weights, i.e., $\lambda_j(i)=1\;$ $\forall i,j$. Therefore, as using constant equal loss weights is also the less computationally expensive approach, we will only present results using this scheme.

\section{Experimental Framework}
\label{sec:framework}


\begin{figure}
  \centering   
  \begin{overpic}[scale=0.5]{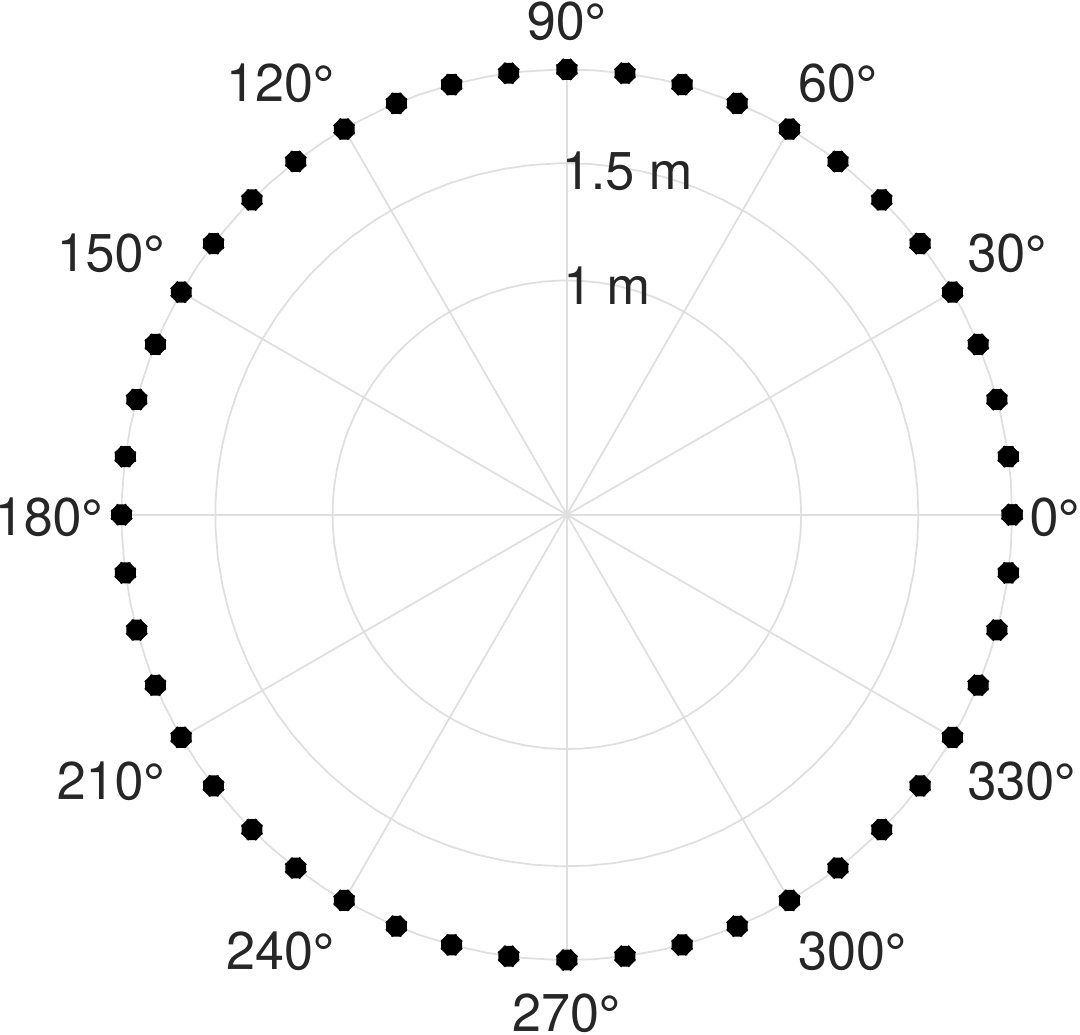}
     \put(44.6,39.3){\includegraphics[scale=0.15]{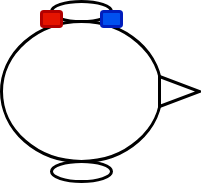}}  
  \end{overpic}
\caption{Every external speaker can be located in one of the 48 equidistantly spaced points (black dots) on a circumference of 1.9 meter radius. An actual person wearing a 2-microphone behind-the-ear hearing aid in her left ear is seated in the center of the circumference. The blue and red dots symbolize the front and rear microphones, respectively, of the hearing aid.}
\label{fig:setup}
\end{figure}

\subsection{Hearing Aid Speech Database}
\label{ssec:database}

The Google Speech Commands Dataset (GSCD) \cite{Warden18} is a speech database comprising 105,829 one second long utterances from a total of 2,618 different speakers. As each utterance contains only one word among a set of 35 possible words, this database is well suited for research on KWS. The GSCD also provides six different background noise files.

To create our speech database from the GSCD, we consider the scenario depicted in Figure \ref{fig:setup}. In a low-reverberation listening room, a circular array of 16 loudspeakers are placed equidistantly spaced around an actual sitting person with a diameter of 3.8 meters at eye-height. Then, 48 head-related transfer functions (HRTFs) are measured at an angular resolution of 7.5 degrees by rotating the chair on which person sits \cite{Moore19}. Here, an HRTF refers to the pair of acoustic transfer functions between the source (loudspeaker) and the front and rear microphones of the left ear hearing aid. Similarly, the own-voice transfer function (OVTF) from the mouth of the person to the microphones of her left ear hearing aid is also measured using a reference microphone placed 2 cm in front of the person's mouth. As measurements were recorded at a sampling rate of 44.1 kHz, the GSCD was upsampled prior to filtering the GSCD signals with the impulse responses.

Around 80\% of the GSCD is reserved for model training, while the validation and test sets span another 10\% each. Speakers do not overlap across sets. For each set, around 75\% of the speakers are randomly selected to simulate that they wear hearing aids (own-voice subset). The rest of speakers (external speaker subset) are used to simulate external speakers and the external speaker angle with respect to the simulated user is randomly chosen from the set of 48 angles (see Figure \ref{fig:setup}) on an utterance basis.

For experimental purposes, we train our models to recognize 10 keywords: ``yes'', ``no'', ``up'', ``down'', ``left'', ``right'', ``on'', ``off'', ``stop'' and ``go''. The remaining 25 words of the GSCD are utilized to populate the \emph{unknown word} class. This class, which is balanced across sets, represents around 10\% of the utterances finally employed.

\subsection{Implementation and Training Issues}
\label{ssec:issues}

Data augmentation is applied during training on an utterance basis by taking into consideration the procedure outlined in \cite{tutorial}. First, a time shift of $u$ ms is applied to the utterance, where $u$ is drawn from the uniform distribution $\mathcal{U}(-100,100)$. Next, with a probability of 0.8, a noise segment is randomly cut from one of the background noise files of the GSCD, scaled by a random factor between 0 and 1, and added to the time-shifted utterance. 30\% of the training data is regenerated at each epoch \cite{Tang18b}.

The multi-task deep residual network was implemented using Keras \cite{chollet2015keras}. Similarly to \cite{Tang18b}, different models were trained for a total of 26 epochs (which is more than enough for convergence) by stochastic gradient descent with a momentum of 0.9. Learning rate and learning rate decay were set to 0.1 and $10^{-5}$, respectively. The minibatch size was of 64 training samples. For both KWS and own-voice/external speaker detection tasks, accuracy (i.e., the ratio between the number of correct predictions and the total number of predictions) was considered as performance metric.

\section{Results}
\label{sec:results}

\begin{table}
  \caption{Summary of the distinguishing features of the different systems that are tested.}
  \label{tab:legend}
  \centering
  \scriptsize
  \setlength{\extrarowheight}{1.5pt}
  \resizebox{0.46\textwidth}{!}{%
\begin{tabular}{l|c|c|c|}
  \cline{2-4}
  & \textbf{Architecture} & \textbf{Training data} & \textbf{Input type} \\ \hline
  \multicolumn{1}{|c|}{Baseline} & \texttt{res15} (KWS only) & Own voice & Front and rear mics \\ \hline
  \multicolumn{1}{|c|}{Front} & Multi-task & Own and external voice & Front mic \\
  \multicolumn{1}{|c|}{Rear} & Multi-task & Own and external voice & Rear mic \\ \hline
  \multicolumn{1}{|c|}{Dual} & Multi-task & Own and external voice & Front and rear mics \\ \hline
  \end{tabular}}
  \vspace{-0.25cm}
\end{table}


\begin{table*}[th]
  \caption{Own-voice/external speaker detection and KWS accuracy results, in percentages, with 95\% confidence intervals.}
  \label{tab:results}
  \centering
  \footnotesize
  \setlength{\extrarowheight}{1pt}
  \begin{tabular}{l|cc|c||c|c|}
  \cline{2-6}
  & \multicolumn{3}{c||}{\textbf{Own-voice/External speaker detection}} & \multicolumn{2}{c|}{\textbf{Keyword spotting}} \\ \cline{2-6}
  & \emph{Own-voice subset} & \emph{External speaker subset} & \emph{Overall} & \emph{Own-voice subset} & \emph{Overall} \\ \hline
  \multicolumn{1}{|c|}{Baseline} & --- & --- & --- & 94.21 $\pm$ 0.39 & 71.87 $\pm$ 0.30 \\ \hline
  \multicolumn{1}{|c|}{Front} & 97.49 $\pm$ 1.02 & 80.38 $\pm$ 5.23 & 93.02 $\pm$ 0.76 & 94.28 $\pm$ 0.37 & 89.48 $\pm$ 0.74 \\
  \multicolumn{1}{|c|}{Rear} & 97.28 $\pm$ 1.08 & 79.03 $\pm$ 5.06 & 92.51 $\pm$ 0.68 & 94.48 $\pm$ 0.25 & 89.29 $\pm$ 0.55 \\ \hline
  \multicolumn{1}{|c|}{Dual} & 99.60 $\pm$ 0.22 & 96.22 $\pm$ 1.61 & 98.72 $\pm$ 0.29 & 94.59 $\pm$ 0.32 & 94.86 $\pm$ 0.39 \\ \hline
  \end{tabular}
\end{table*}

We test the multi-task architecture by making use of the dual-microphone signal (\emph{Dual}) from the hearing assistive device and compare it with using the single-microphone signal from the front (\emph{Front}) and rear (\emph{Rear}) microphones, respectively. To assess the KWS performance of existing systems, which do not take the potential presence of external speakers into account, we test the original \texttt{res15} (i.e., with no own-voice/external speaker detection output) using the dual-microphone signal as input (Baseline). For Baseline model training, only own voice ---and not external speaker data--- is employed. For the sake of clarity, Table \ref{tab:legend} summarizes the distinguishing features of the different systems that are tested.

\subsection{Own-Voice/External Speaker Detection Results}

The left part of Table \ref{tab:results} presents the own-voice/external speaker detection accuracy results\footnote{These results were obtained by making use of a sigmoid decision threshold of 0.5, which is equivalent to detecting the most likely class.}, in percentages, with 95\% confidence intervals across 10 different networks trained with different random model initialization. Overall (i.e., over the whole test set) own-voice/external speaker detection accuracy results are also broken down by accuracies measured separately on the own-voice and external speaker subsets of the test set.

\begin{figure}[t]
  \centering
  \includegraphics[width=0.9\linewidth]{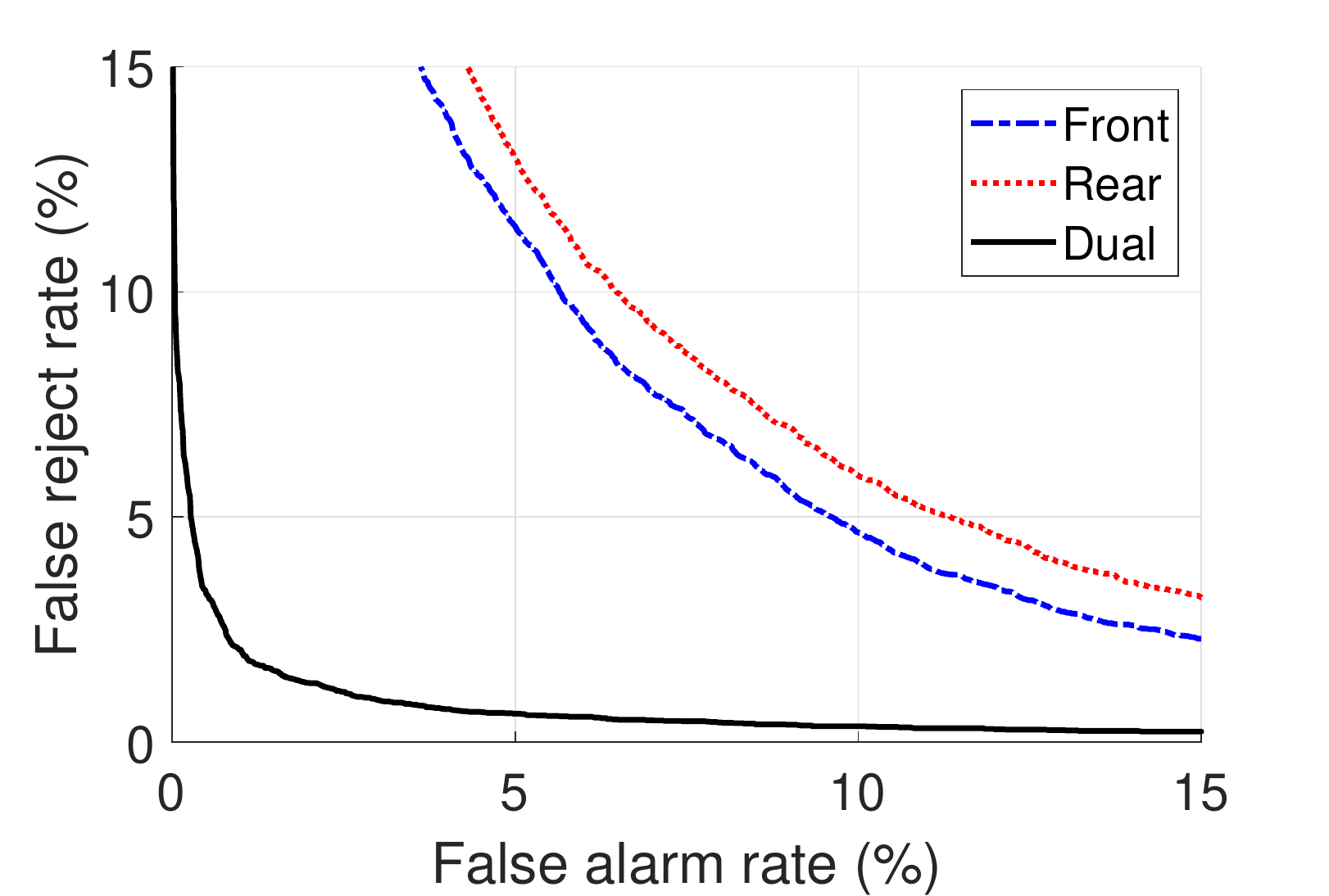}
  \caption{Detection error trade-off curves for own-voice/external speaker detection.}
  \label{fig:roc}
  \vspace{0.1cm}
\end{figure}

From the left part of Table \ref{tab:results}, we can see that own-voice detection is accomplished with a high accuracy (between 97\% and 99\%) by exploiting either the front or rear microphone or both of them simultaneously. Using the dual-microphone signal helps for external speaker detection by clearly outperforming the single-channel strategy. These results are supported by the detection error trade-off curves for own-voice/external speaker detection in Figure \ref{fig:roc}. In this figure, pairs of false alarm rate and false reject rate values are plotted as a function of the sigmoid decision threshold (which is swept from 0 to 1). The smaller the area under the curve, the better a system is. Exploiting either the front or rear microphone works similarly with no statistically significant differences, and using the dual-microphone signal provides the best detection performance by far.

\begin{figure}[t]
  \centering   
  \begin{overpic}[width=0.75\linewidth]{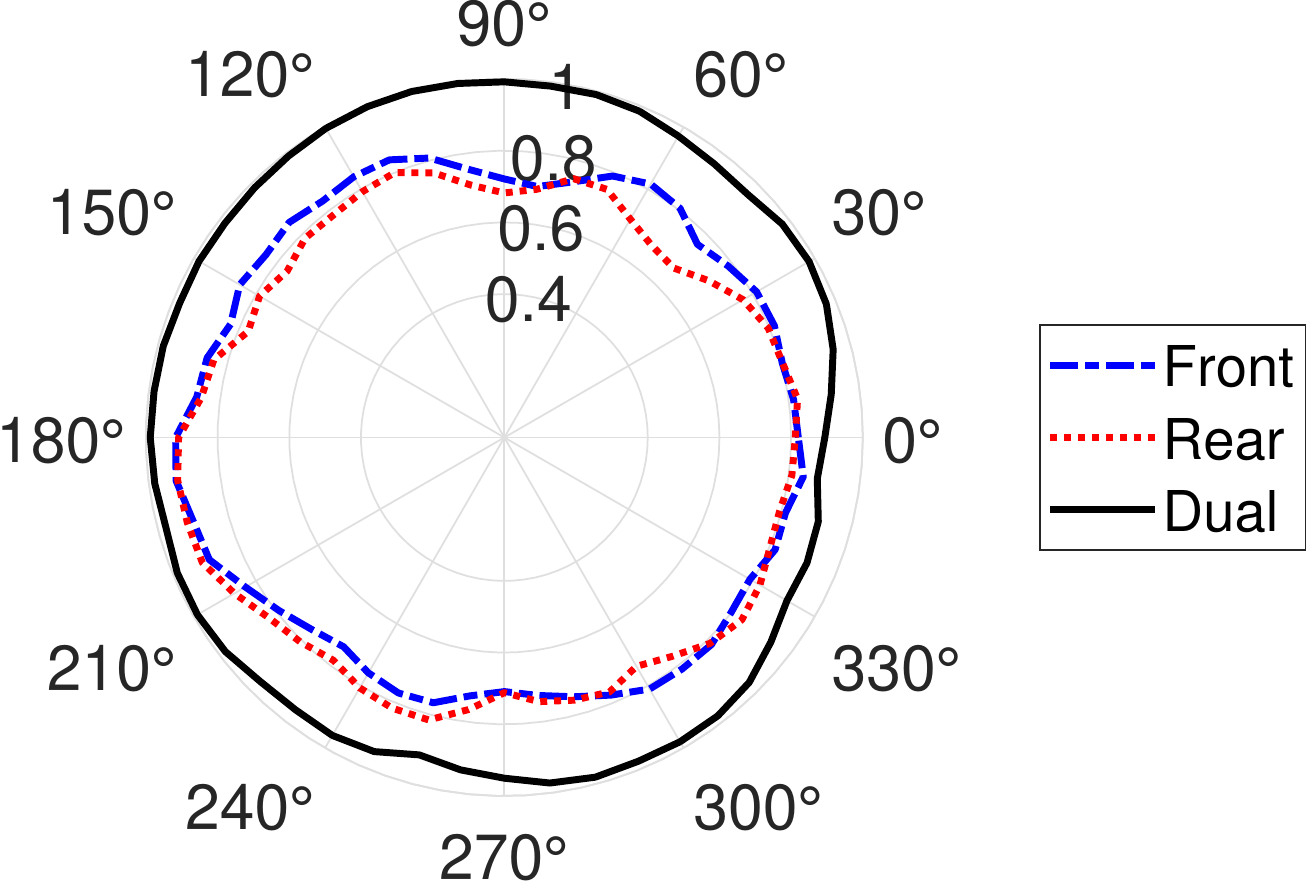}
     \put(31.3,25.9){\includegraphics[scale=0.15]{AidsPNG}}  
  \end{overpic}
  \caption{Normalized external speaker detection accuracy as a function of the external speaker angle with respect to the user of the hearing assistive device. The users' head is centered in the origin and faces towards 0°.}
  \label{fig:det}
\end{figure}

Figure \ref{fig:det} depicts the normalized external speaker detection accuracy as a function of the external speaker angle with respect to the hearing aid user. We hypothesize that the particular contour of these curves may be explained by the characteristics of the OVTF and HRTFs. This hypothesis is supported by an analysis of the OVTF and HRTFs, which showed that they are more similar (in terms of MFCC Euclidean distance) at angles where we see a relative drop in performance (e.g., for \emph{Dual}, at frontal and shadow side [$\sim$250°] angles). In other words, these similarities would make an external speaker less distinguishable from the user, thereby yielding a relative worsening in terms of external speaker detection accuracy.

\subsection{Keyword Spotting Results}

The right part of Table \ref{tab:results} presents the KWS accuracy results. For overall KWS accuracy computation, own-voice/external speaker detection is taken into account. Thus, correct decisions are made in the following two cases: \emph{1)} when the hearing aid user triggers the system as a result of right keyword recognition, and \emph{2)} otherwise (i.e., when the user speaks but it is not a keyword, or an external speaker utters a keyword or something different) the KWS system is not triggered. To understand the KWS performance degradation due to external speakers, KWS results on the own-voice subset of the test set are also reported.

The right part of Table \ref{tab:results} reveals the impact of own-voice/external speaker detection on KWS accuracy. Baseline results justify the need for effective own-voice/external speaker detection, as we propose. In other words, while the \texttt{res15} architecture achieves a high own-voice KWS accuracy, its performance drops significantly in the presence of external speakers. This performance degradation is clearly alleviated by using the multi-task architecture along with either the front or rear microphone signals. As expected, the best KWS accuracy results are obtained by jointly exploiting the multi-task learning scheme and the dual-microphone signal. This approach ($\sim$94.86\% acc.) achieves an overall KWS accuracy relative improvement of around 32\% with respect to Baseline ($\sim$71.87\% acc.) and, more importantly, there is no drop between own-voice and overall KWS accuracies.

\section{Conclusions}
\label{sec:conclusions}

In this paper we have proposed a multi-task learning strategy to carry out KWS for hearing assistive devices that is robust to external speakers. This robustness is important for practical applications like the one assessed here, where our approach has been able to significantly outperform a state-of-the-art small-footprint KWS system. Furthermore, this has been achieved with a negligible increase in the number of parameters of the model.


\section{Acknowledgements}

This work was supported, in part, by the Oticon Foundation.

\bibliographystyle{IEEEtran}

\bibliography{mybib}

\end{document}